\documentclass[review]{elsarticle}

\graphicspath{ {./figures/} }
\usepackage{amsmath} 
\usepackage{hyperref}
\usepackage{float}
\usepackage{verbatim} %comments
\usepackage{apalike}
\restylefloat{figure} 
\floatstyle{plaintop} %table caption at top
\restylefloat{table}

\journal{Expert Systems with Applications}

\biboptions{authoryear}

\begin{document}
\begin{frontmatter}

%The title page must contain the title of the paper and the full name/full affiliation with country/e-mail address for each author and co-author of the manuscript. Please make sure you have included all elements listed below with your manuscript submission.

% \begin{titlepage}
% \begin{center}
% \vspace*{1cm}

% \textbf{ \large Enhancing Stock Market Information Retrieval with Retrieval-Augmented Generation}

% \vspace{1.5cm}

% % Author names and affiliations
% Huan Phong Thai$^{a}$ (huantpce180685@fpt.edu.vn), Thu Anh Ngoc Le$^a$ (thulnace180615@fpt.edu.vn), Nhu Huyenh Thi Le$^a$ (nhulthca182007@fpt.edu.vn), Nhan Huu Tran(nhanthce181655@fpt.edu.vn), Lenh Cong Phan Pham, Hong Nhung Nguyen$^b$ (nhungnguyen.uet@gmail.com) \\

% \hspace{10pt}

% \begin{flushleft}
% \small  
% $^a$  Department of Software Engineering, FPT University, Can Tho, Viet Nam \\
% $^b$  Department of Artificial Intelligence, Gachon University, Seongname-si, South Korea \\

% \begin{comment}
% Clearly indicate who will handle correspondence at all stages of refereeing and publication, also post-publication. Ensure that phone numbers (with country and area code) are provided in addition to the e-mail address and the complete postal address. Contact details must be kept up to date by the corresponding author.
% \end{comment}

% \vspace{1cm}
% \textbf{Corresponding author at: Hong Nhung Nguyen, Gachon University, South Korea.} \\

% Tel: (+010)8129-2505 \\
% Email: nhungnguyen.uet@gmail.com

% \end{flushleft}        
% \end{center}
% \end{titlepage}

\title{Enhancing Stock Market Information Retrieval with Retrieval-Augmented Generation}

\author[label1]{Huan Phong Thai}
\ead{huantpce180685@fpt.edu.vn}

\author[label1]{Thu Anh Ngoc Le}
\ead{thulnace180615@fpt.edu.vn}

\author[label1]{Nhu Huynh Thi Le}
\ead{nhulthca182007@fpt.edu.vn}

\author[label1]{Nhan Huu Tran}
\ead{nhanthce181655@fpt.edu.vn}

\author[label1]{Lenh Cong Phan Pham}
\ead{lenhppcce180059@fpt.edu.vn}

\author[label1]{Hong Nhung Nguyen \corref{cor1}}
\ead{nhungnguyen.uet@gmail.com}

\cortext[cor1]{Corresponding author.}

\address[label1]{Department of Software Engineering, FPT University, Can Tho, Viet Nam}
% \address[label2]{Department of Artificial Intelligence, Gachon University, Republic of Korea}

\begin{abstract}
Natural Language Processing (NLP) has been widely applied in financial domains, enabling automated analysis and retrieval of stock market information. However, traditional NLP-based systems often struggle with maintaining contextual accuracy and generating relevant responses in dynamic financial environments. To address these challenges, Retrieval-Augmented Generation (RAG) models have emerged as a powerful solution, combining retrieval-based methods with generative capabilities to enhance response quality. In this study, we present a Retrieval-Augmented Generation (RAG) system designed to generate responses related to stock market information. The system integrates retrieval-based and generative approaches to enhance the accuracy and relevance of financial text generation. To evaluate its effectiveness, we employ the ROUGE metric, specifically ROUGE-1, ROUGE-2, and ROUGE-L, to measure lexical and structural similarity between generated and reference texts. Experimental results on a dataset of 100 stock market-related queries demonstrate high ROUGE scores, indicating strong alignment with reference documents. Additionally, we assess latency to evaluate the system’s responsiveness, achieving an average response time of 1.16 seconds. These findings confirm that our RAG model is capable of providing accurate and timely stock market insights, making it a promising approach for financial information retrieval.
\end{abstract}


\end{frontmatter}

\section{Introduction}
\label{introduction}

Large Language Models (LLMs) \cite{b1} \cite{b2} have made remarkable progress in recent years, establishing themselves as a fundamental component in various natural language processing (NLP) applications. Their ability to understand and generate human-like text has led to significant advancements in tasks such as machine translation, sentiment analysis, question answering, and text summarization. As a result, LLMs have been widely adopted across multiple domains, including education \cite{b3}, healthcare \cite{b4}, finance \cite{b5}, code writing \cite{b6}, cybersecurity \cite{b7}, the labor market \cite{b8}, and legal aspects \cite{b9}, where they enhance automation, improve decision-making, and enable more natural human-computer interactions. One of the most promising approaches in this domain is Retrieval-Augmented Generation (RAG), which combines the strengths of retrieval-based and generative models to enhance the accuracy and reliability of generated content. 

Despite the impressive capabilities of existing LLMs, their development is still constrained by several challenges. One of the most significant issues is the hallucination problem \cite{b10}, where models generate responses that are fluent and coherent but factually inaccurate. Another major limitation is the difficulty in updating their knowledge. Since LLMs store information within their internal memory \cite{b11}, keeping them up to date requires retraining or fine-tuning with new data—an expensive and resource-intensive process. Additionally, general-purpose LLMs often lack expertise in specialized domains \cite{b12} \cite{b13}, making them less effective for domain-specific tasks. Developing LLMs tailored to a particular field requires extensive effort in dataset collection and model training, further increasing the complexity of implementation. This makes the deployment of systems for specific tasks costly in terms of both time and resources. Furthermore, collecting and processing training data for each specialized domain can present challenges such as a lack of high-quality data, security requirements, and the complexity of fine-tuning the model to fit the specific context.

Stock markets are highly dynamic, influenced by a wide range of factors, including economic indicators, corporate financial reports, geopolitical events, and investor sentiment. Efficiently retrieving relevant financial information and generating accurate insights is crucial for investors, analysts, and decision-makers. Traditional NLP models often struggle with the complexity and ever-changing nature of financial data, making it challenging to provide real-time, contextually accurate, and explainable outputs. To solve this problem, RAG presents a potential solution by integrating retrieval mechanisms with generative capabilities, allowing models to dynamically fetch the most relevant financial data and generate informed responses.

\begin{figure*}[htbp]
    \centering
    \includegraphics[width=\textwidth]{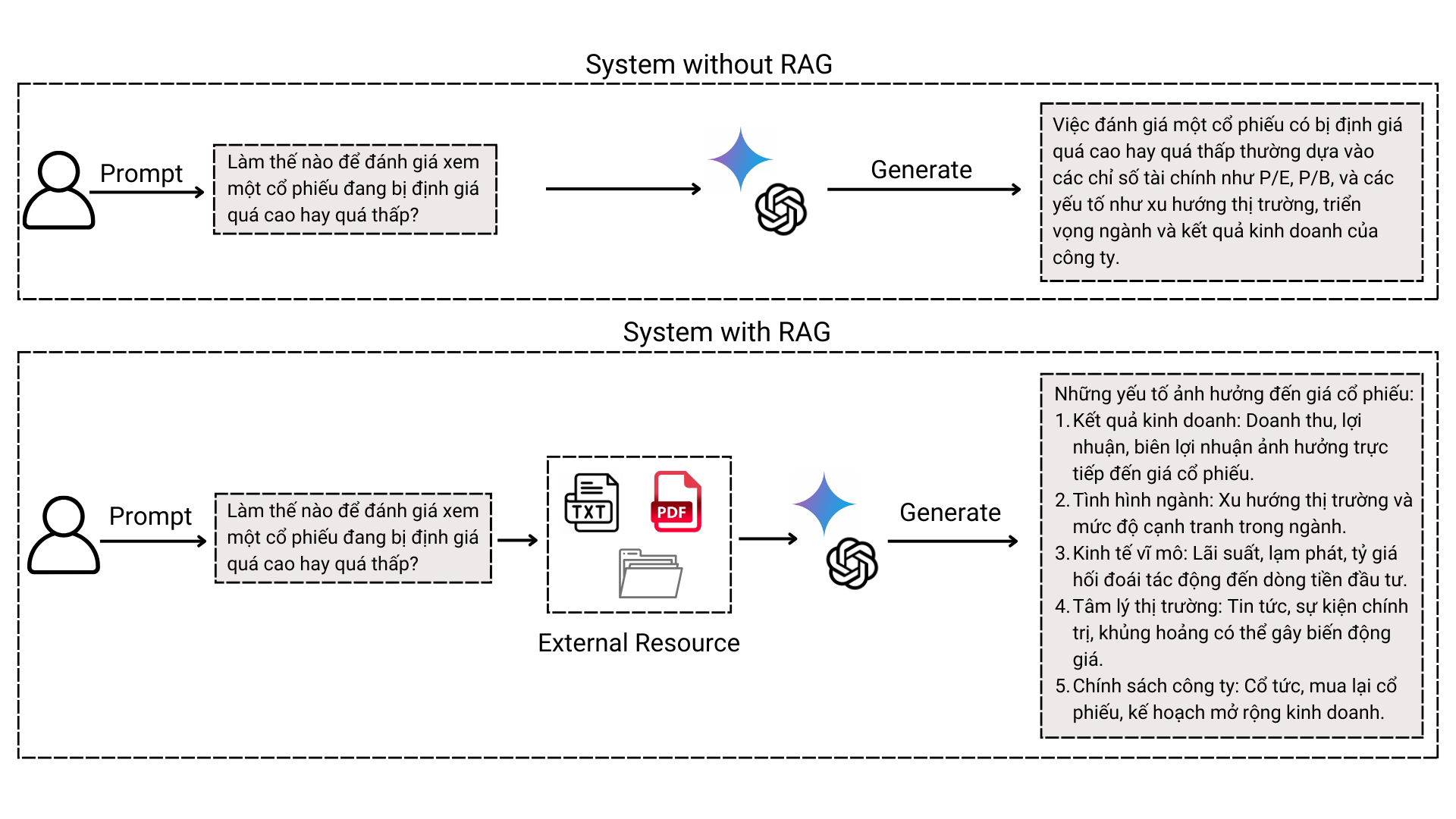}
    \caption{System without RAG and System with RAG}
    \label{fig1}
\end{figure*}

Several recent studies \cite{b14} \cite{b15} have proposed leveraging external knowledge bases to enhance the ability of LLMs to generate better responses, a technique known as Retrieval-Augmented Generation (RAG). By supplying LLMs with retrieved relevant factual information, the hallucination problem can be alleviated to some extent. Moreover, the knowledge update issue can also be addressed by updating the external knowledge database, allowing LLMs to incorporate up-to-date information. RAG can also transform a general LLM into a domain-specific LLM by constructing and utilizing a specialized knowledge database. Therefore, RAG plays a crucial role in enhancing the functionality of LLMs, making them more accurate, knowledgeable, and reliable across a wide range of applications \cite{b16}. Fig. \ref{fig1} illustrates the fundamental differences in workflow between a system without RAG and a system enhanced with RAG, highlighting the impact of retrieval-augmented knowledge integration.

This study aims to explore the effectiveness of RAG models in processing stock market information. Specifically, we investigate how RAG can enhance the retrieval of information and the generation of responses regarding stock market news published on VnExpress. By leveraging external financial knowledge sources, RAG models may enhance decision-making processes, reduce misinformation, and offer more comprehensive market summaries. In summary, the structure of this paper is as follows: Section 2 reviews related work in financial NLP and RAG applications. Section 3 describes the proposed methodology, including data sources, model architecture, and evaluation metrics. Section 4 presents experimental results and discussions. Finally, Section 5 concludes the paper with key findings and potential future research directions.

\section{Related Work}
\label{related work}
\subsection{NLP in Stock Market}

Financial NLP has been extensively used to analyze stock market data, predict trends, and support investment decisions. Several studies have explored the effectiveness of NLP models in processing financial iinformation. Financial NLP has been extensively used to analyze stock market data, predict trends, and support investment decisions. Several studies have explored the effectiveness of NLP models in processing financial information. Akita et al. \cite{b17} utilized Long Short-Term Memory (LSTM) networks to predict stock prices by combining textual and numerical data from the Nikkei newspaper. Their research highlighted the importance of cross-company event correlation in financial forecasting, demonstrating how news about one company could impact the stock prices of others within the same industry. Similarly, Sidra and Jaydip \cite{b18} explored multiple machine learning and deep learning techniques for weekly stock market predictions, integrating sentiment analysis from social media with historical stock prices. Their findings showed that public sentiment plays a crucial role in improving stock movement predictions.

Another notable study proposed an LSTM-based forecasting model that incorporated sentiment analysis to establish correlations between stock values and news sentiment. The results demonstrated that using NLP for feature extraction and LSTM for time-series training significantly improved prediction accuracy, reducing the mean square error to 0.062 \cite{b19}. Additionally, "Smart Trader," an AI-powered chatbot designed for the Saudi Tadawul market, showcased the practical applications of NLP and machine learning in stock trading. By leveraging real-time market data and user preferences, the chatbot provided personalized investment recommendations, filling a gap in the market for AI-driven financial advisory services \cite{b20}.

\subsection{Retrieval-Augmented Generation (RAG) in NLP}

Retrieval-Augmented Generation (RAG) is a hybrid approach that enhances language models by integrating external knowledge retrieval mechanisms. Instead of relying solely on parametric memory, RAG retrieves relevant documents or facts from external sources and conditions its responses based on retrieved content. This method is particularly useful for knowledge-intensive NLP tasks such as open-domain question answering and fact-based text generation \cite{b21}. By dynamically incorporating relevant information at inference time, RAG models can mitigate issues like hallucination and outdated knowledge, which are common challenges in traditional language models.

The architecture of RAG consists of two main components: a retriever module and a generator module. The retriever module searches for relevant external knowledge, while the generator module synthesizes a response based on the retrieved content. The retriever module usually comprises three components: an encoder for encoding inputs into embeddings, an efficient indexing system that supports approximate nearest neighbor search, and a datastore for storing external knowledge in the form of key-value pairs \cite{b16}. Beside, Retrieval-augmented (RA) generators and default generators are the two primary types of generator modules. Default generators encompass most pre-trained or fine-tuned large language models, including Mistral models \cite{b22}, Gemini-series models \cite{b1} \cite{b23}, and GPT-series models \cite{b2} \cite{b24}. In contrast, RA generators integrate retrieval mechanisms to enhance response generation, incorporating modules like Enc-Dec \cite{b25} and RETRO \cite{b26}. These generators refine their outputs by leveraging retrieved knowledge, allowing them to generate more contextually informed responses or predictions.

Recent studies have demonstrated the effectiveness of RAG in improving factual accuracy and reducing misinformation in AI-generated text. For example, RAG models have been used to enhance chatbot responses by retrieving real-time information from external databases \cite{b21}. Additionally, advances in retrieval mechanisms, such as dense passage retrieval and contrastive learning, have further improved the quality of retrieved content, leading to more coherent and contextually relevant responses.
\subsection{RAG in Financial and Stock Market Applications}
Retrieval-Augmented Generation (RAG) has been increasingly applied in financial domains to enhance decision-making by integrating real-time knowledge retrieval with language model capabilities. In market sentiment analysis, fine-tuned models like FinBERT within a RAG pipeline have been used to analyze financial news and predict stock price movements, though sentiment alone shows limited predictive power \cite{b27}. Similarly, RAG improves financial analysis by addressing challenges such as lack of interpretability in ML models and hallucinations in LLMs. The AlphaFin dataset and Stock-Chain framework have demonstrated the effectiveness of integrating retrieval mechanisms for financial reasoning tasks \cite{b28}.

Beyond sentiment analysis, RAG plays a crucial role in multi-modal financial analytics. The FinTMMBench benchmark evaluates temporal-aware RAG systems that process diverse financial data sources, such as stock prices, news, and technical charts, enabling more comprehensive trend analysis and event detection \cite{b29}. Additionally, in financial risk assessment, RAG enhances LLM performance in audit report analysis by improving faithfulness and context relevance. Experiments show that models like LLaMa3.1 and Gemini-1.5-flash significantly improve financial risk evaluations by extracting detailed insights from reports \cite{30}. These applications highlight the potential of RAG in financial markets, bridging the gap between real-time information retrieval and AI-generated financial insights.

\section{Data Preparation}
\label{data preparation}
\subsection{Data Collection}

In this study, we collected data from VnExpress, a well-known Vietnamese online news platform, to analyze stock market trends. The dataset comprises 30 news articles related to the stock market, which were crawled from VnExpress. These articles cover various aspects of the stock market, including trading volume, stock price fluctuations, corporate financial performance, and macroeconomic factors influencing market trends.

\begin{table}[htbp]
\caption{Number of Main Documents}
\begin{center}
\label{tab:document_categories}
\begin{tabular}{|l|c|}
\hline
\textbf{Category} & \textbf{Number} \\
\hline
Market Trends \& Index Performance & 8 \\
\hline
Corporate Stocks \& Sectors & 9 \\
\hline
Foreign Investment \& Liquidity & 7 \\
\hline
Regulatory \& Corporate Actions & 6 \\
\hline
\textbf{Total} & \textbf{30} \\
\hline
\end{tabular}
\end{center}
\end{table}

Each category represents a key dimension of the stock market. Market Trends \& Index Performance tracks overall market movement and index fluctuations. Corporate Stocks \& Sectors highlights individual stock performances and sector-wide trends. Foreign Investment \& Liquidity focuses on the role of international capital in the Vietnamese stock market. Lastly, Regulatory \& Corporate Actions examines government policies, financial regulations, and corporate decisions that influence market behavior.

\begin{figure*}[htbp]
    \centering
    \includegraphics[width=0.75\textwidth]{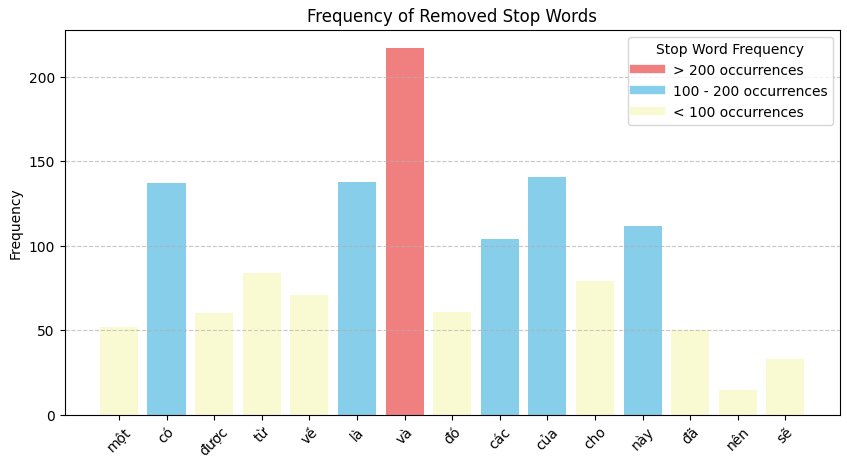}
    \caption{Frequency of Removed Stop Words with Categorized Colors}
    \label{stop_words}
\end{figure*}

This categorization helps in understanding the key trends and themes in the stock market as reported by VnExpress. The data will be utilized for further analysis to develop a Retrieval-Augmented Generation (RAG) system that enhances stock market news comprehension and decision-making support.

\subsection{Data Preprocessing}

Stop words are commonly used words in a language that carry minimal semantic meaning, such as conjunctions, prepositions, and auxiliary verbs. Removing stop words is a crucial step in text preprocessing as it helps reduce dimensionality, improve computational efficiency, and enhance the relevance of text-based features for downstream natural language processing (NLP) tasks. Fig \ref{stop_words} illustrates the frequency of removed stop words in the documents.

In this study, we implemented a stop word removal process specifically for Vietnamese text. Given the unique characteristics of the Vietnamese language, where words are often multi-syllabic and require tokenization before processing, we utilized the Underthesea library for word segmentation. The preprocessing pipeline consists of several key steps.

First, the input text was normalized by stripping leading and trailing whitespaces. Additionally, all URLs and special characters were removed using regular expressions to eliminate noise in the data. To ensure uniformity, all text was converted to lowercase, which helps in minimizing variations due to capitalization.

Next, since Vietnamese is not naturally space-separated like English, we applied word\_tokenize from the Underthesea library to segment text into meaningful tokens. Following tokenization, we curated a list of common Vietnamese stop words and removed them from the tokenized text. This step reduces redundancy and ensures that only semantically meaningful words are retained for further analysis.

After performing these preprocessing steps, the cleaned text was stored for subsequent stages of the information retrieval and retrieval-augmented generation (RAG) pipeline. The removal of stop words significantly enhances the quality of text embeddings by focusing on content-bearing words while discarding non-informative tokens. This refinement is particularly beneficial for improving the performance of semantic similarity searches and document retrieval tasks.

\section{Proposed Methodology}
\label{proposed methodology}
\begin{figure*}[htbp]
    \centering
    \includegraphics[width=0.7\textwidth]{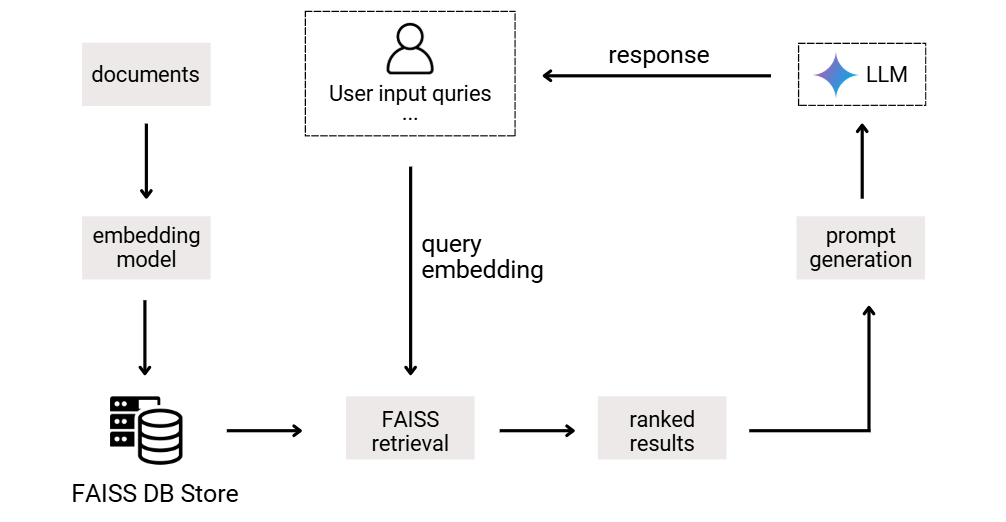}
    \caption{Retrieval-Augmented Generation (RAG) System Architecture}
    \label{fig2}
\end{figure*}

This section presents the architecture of the proposed Retrieval-Augmented Generation (RAG) system, which enhances large language model (LLM) responses by integrating an external knowledge retrieval mechanism. The system consists of three main components: text embedding, information retrieval, and response generation. Fig. \ref{fig2} illustrates the overall workflow of the RAG system, depicting how queries are processed through embedding, information retrieval, and response generation.

\subsection{Text Embedding}

\begin{figure*}[htbp]
    \centering
    \includegraphics[width=0.7\textwidth]{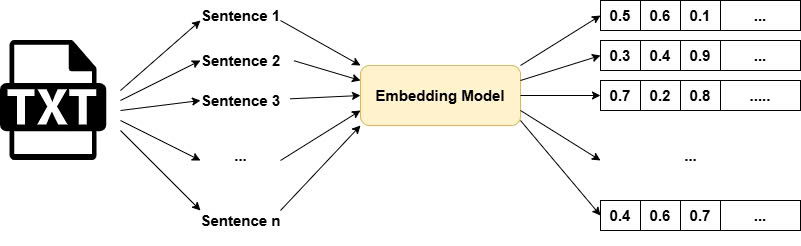}
    \caption{Text Embedding Process: Converting Raw Text into Numerical Representations for Efficient Retrieval}
    \label{fig3}
\end{figure*}

The embedding process converts textual documents into numerical vectors using a pretrained embedding model. Fig \ref{fig3} illustrates the process of embedding textual data into numerical vector representations. As mentioned above, the documents consist of stock market news articles crawled from VnExpress, one of Vietnam’s most popular online newspapers. VnExpress provides up-to-date financial news, market analysis, and expert opinions, making it a valuable data source for financial insights.

Specifically, we employ the LaBSE (Language-agnostic BERT Sentence Embedding) \cite{b31} model from the SentenceTransformers library. This model is chosen for its multilingual capabilities, allowing it to effectively encode text in various languages. Notably, LaBSE provides strong support for the Vietnamese language, ensuring high-quality semantic representations of Vietnamese financial news.

Documents are first preprocessed and tokenized using Jieba \cite{b32}, a widely used text segmentation tool primarily designed for Chinese. Despite its origin, jieba has been adapted for Vietnamese and is used here to segment sentences into meaningful tokens before embedding. Proper tokenization is crucial for accurately representing text semantics.

The embeddings are then computed and stored in a FAISS (Facebook AI Similarity Search) \cite{b33} database, which enables efficient nearest-neighbor retrieval. The generated embeddings form a high-dimensional vector space that facilitates rapid similarity comparisons. Once all textual data has been transformed into vector representations, they are systematically stored in the FAISS DB store for efficient retrieval during the next processing stages.

\subsection{Information Retrieval}

Given a user query, the retrieval process aims to fetch the most relevant documents from the FAISS database. The retrieval mechanism follows a hybrid approach that integrates FAISS-based vector similarity search with BM25 (Best Matching 25) \cite{b34} ranking. This combination leverages both semantic similarity and term frequency-based relevance to enhance the accuracy of retrieved documents.

To perform vector-based retrieval, the input query is first embedded using the LaBSE model, ensuring a dense numerical representation in the same vector space as the stored documents. This query embedding is then compared against the stored document embeddings using FAISS. FAISS efficiently retrieves the top-k nearest neighbors by measuring the L2 distance (Euclidean distance) between the query and stored vectors. The L2 distance between two vectors \textbf{q} (query) and \textbf{d} (document) in an n-dimensional space is computed as follows:

\begin{equation}
d(\mathbf{q}, \mathbf{d}) = \sqrt{\sum_{i=1}^{n} (q_i - d_i)^2}
\end{equation}

where: \( q_i \) and \( d_i \) represent the individual components of the query and document embeddings, respectively.  
A smaller L2 distance indicates higher similarity between the query and document.

After retrieving an initial set of candidate documents from FAISS, a BM25 ranking step is applied to further refine the results. BM25 is a TF-IDF-based retrieval model that scores documents based on their word frequency, document length normalization, and inverse document frequency (IDF). This step ensures that retrieved documents are not only semantically similar but also contextually relevant based on keyword occurrences. The BM25 score for a document \( D \) given a query \( Q \) is computed as:

\begin{equation}
BM25(D, Q) = \sum_{t \in Q} IDF(t) \cdot \frac{f(t, D) \cdot (k_1 + 1)}{f(t, D) + k_1 \cdot \left(1 - b + b \cdot \frac{|D|}{\text{avgD}}\right)}
\end{equation}

where:
\begin{itemize}
    \item \( f(t, D) \) is the term frequency of term \( t \) in document \( D \).
    \item \( |D| \) is the length of document \( D \) (total number of words in \( D \)).
    \item \( \text{avgD} \) is the average document length in the corpus.
    \item \( k_1 \) is a tuning parameter controlling term frequency saturation.
    \item \( b \) is a tuning parameter controlling document length normalization.
\end{itemize}

The inverse document frequency (IDF) component is defined as:

\begin{equation}
IDF(t) = \log \frac{N - n(t) + 0.5}{n(t) + 0.5}
\end{equation}

where:
\begin{itemize}
    \item \( N \) is the total number of documents in the corpus.
    \item \( n(t) \) is the number of documents containing term \( t \).
\end{itemize}
\vspace{0.5cm}
The BM25 ranking model is particularly effective in handling keyword-based relevance, especially for queries where specific terms are important. By applying BM25 after FAISS retrieval, we re-rank the initially retrieved documents to prioritize those that best match the user’s intent.

This hybrid approach ensures that the final set of documents includes both semantically related and keyword-relevant results, optimizing the retrieval process for accuracy and contextual relevance.

\subsection{Information Generation}

The final stage of the system focuses on generating a response based on the retrieved documents. We utilize the Gemini model, specifically Gemini-1.5-Flash \cite{b22} \cite{b35}, due to its optimized performance for real-time applications, lower latency, and ability to process long-context inputs efficiently. This choice ensures that the model can effectively integrate retrieved knowledge while maintaining responsiveness.

To generate a well-informed response, a structured prompt is carefully crafted by incorporating the retrieved documents as contextual information. The prompt is formatted to include a clear instruction, the user’s query, and the most relevant retrieved texts, ensuring that the model leverages external knowledge effectively.

Once the structured prompt is prepared, Gemini-1.5-Flash processes it and generates a response that aligns factually with the retrieved information. The model does this by grounding its output in the provided texts, minimizing hallucinations, and maintaining consistency with real-world stocket market informations.

By leveraging this retrieval-augmented approach, the system significantly enhances the quality of responses, ensuring that they are not only coherent and contextually aware but also well-supported by factual knowledge.

\section{Experimental Result}

\subsection{Evaluation Metrics}

To assess the performance of our Retrieval-Augmented Generation (RAG) system in generating responses related to stock market information, we employ ROUGE (Recall-Oriented Understudy for Gisting Evaluation) Score as the primary evaluation metric. ROUGE is a widely used metric for evaluating text summarization and generation tasks by comparing the overlap of n-grams, word sequences, and longest common subsequences between the generated text and reference answers.

We calculate ROUGE-1, ROUGE-2, and ROUGE-L scores, which are defined as follows:

\begin{itemize}
    \item \textbf{ROUGE-1} measures the overlap of unigrams (single words) between the generated text and the reference text:
    \begin{equation}
        \text{ROUGE-1} = \frac{\sum_{w \in W} \min(\text{count}_{\text{gen}}(w), \text{count}_{\text{ref}}(w))}{\sum_{w \in W} \text{count}_{\text{ref}}(w)}
    \end{equation}

    \item \textbf{ROUGE-2} extends this to bigrams (two consecutive words), capturing more contextual meaning:
    \begin{equation}
        \text{ROUGE-2} = \frac{\sum_{b \in B} \min(\text{count}_{\text{gen}}(b), \text{count}_{\text{ref}}(b))}{\sum_{b \in B} \text{count}_{\text{ref}}(b)}
    \end{equation}

    \item \textbf{ROUGE-L} evaluates the longest common subsequence (LCS) between the generated and reference text, which captures sentence-level structure and fluency:
    \begin{equation}
        \text{ROUGE-L} = \frac{LCS(\text{gen}, \text{ref})}{\text{length}(\text{ref})}
    \end{equation}
\end{itemize}

Where:
\begin{itemize}
    \item $\text{count}_{\text{gen}}(w)$ and $\text{count}_{\text{ref}}(w)$ are the counts of word \(w\) in the generated and reference texts, respectively.
    \item $W$ and $B$ are the sets of all unigrams and bigrams in the reference text.
    \item $LCS(\text{gen}, \text{ref})$ is the length of the longest common subsequence between the generated and reference text.
    \item $\text{length}(\text{ref})$ is the length of the reference text.
\end{itemize}

In addition to ROUGE scores, we measure Latency, which quantifies the system's response time. Latency is defined as the time elapsed from the moment a query is submitted to the system until a response is generated. This metric is critical for assessing the real-time usability of our RAG system, particularly in financial applications where timely information retrieval is crucial.

By combining accuracy-based ROUGE metrics with latency measurement, we obtain a comprehensive evaluation of both the quality and efficiency of our system's responses.

\subsection{Retrieval-Augmented Generation (RAG) Results}

To evaluate the effectiveness of our RAG system, we computed ROUGE scores based on a sample of 100 generated responses derived from stock market-related documents. The average ROUGE scores obtained are presented in Tab \ref{tab:rouge_scores}

\begin{table}[htbp]
\caption{ROUGE Score Results}
\begin{center}
\label{tab:rouge_scores}
\begin{tabular}{|l|c|c|c|}
\hline
\textbf{ROUGE Metric} & \textbf{Precision} & \textbf{Recall} & \textbf{F-measure} \\
\hline
ROUGE-1 & 0.9616 & 0.9597 & 0.9555 \\
\hline
ROUGE-2 & 0.9374 & 0.9356 & 0.9314 \\
\hline
ROUGE-L & 0.9496 & 0.9478 & 0.9436 \\
\hline
\end{tabular}
\end{center}
\end{table}

These results indicate that the system achieves high lexical and structural similarity between the generated responses and reference texts, demonstrating the capability of the RAG model in preserving key stock market information from the original documents. The high precision and recall values across all ROUGE metrics suggest that the system generates responses that closely align with the reference texts, ensuring both completeness and correctness in retrieving stock market-related data.

Specifically, the ROUGE-1 score of 0.9555 (F-measure) reflects the system's ability to capture important individual keywords relevant to stock market discussions, ensuring that essential financial terms and concepts are accurately represented in the generated responses. Meanwhile, the ROUGE-2 score of 0.9314 indicates that the model effectively retains key bi-gram (two-word phrase) structures, which are crucial for maintaining contextual accuracy and conveying financial insights with coherence.

Moreover, the strong ROUGE-L score of 0.9436 highlights that the system maintains sentence structure and fluency, further reinforcing its reliability for financial text generation tasks. These consistently high scores across all three ROUGE metrics confirm the robustness of our approach in providing high-quality, contextually accurate, and well-structured responses for stock market information retrieval.

Apart from accuracy, latency performance was also measured to assess system efficiency. Over 100 queries, the latency statistics are as follows Tab \ref{tab:latency_results}:

\begin{table}[htbp]
\caption{Latency Results}
\begin{center}
\label{tab:latency_results}
\begin{tabular}{|l|c|}
\hline
\textbf{Latency Metric} & \textbf{Time (seconds)} \\
\hline
Average Latency & 1.1626 \\
\hline
Minimum Latency & 0.917 \\
\hline
Maximum Latency & 2.455 \\
\hline
\end{tabular}
\end{center}
\end{table}

These latency results suggest that the system provides responses within an acceptable time frame, balancing both accuracy and efficiency. The combination of high ROUGE scores and low latency values demonstrates the effectiveness of our RAG model for financial information retrieval. The relatively low response time further supports its practical usability in real-world applications where timely access to accurate stock market information is critical.

\section{Conclusion}

In this study, we evaluated the effectiveness of a Retrieval-Augmented Generation (RAG) system in generating responses related to stock market information. By utilizing the ROUGE metric, we assessed the lexical and structural similarity between the generated responses and reference texts. The high ROUGE-1, ROUGE-2, and ROUGE-L scores demonstrate that the system is capable of accurately retrieving and synthesizing financial data while maintaining coherence and fluency in the responses. Additionally, our analysis of latency indicates that the system operates within an acceptable response time, ensuring efficient query processing for real-time financial applications. The combination of high accuracy and low latency confirms that the proposed RAG model is a viable approach for handling stock market-related queries, providing reliable and timely information. Future work will focus on further improving the model’s ability to handle complex financial queries, integrating external knowledge sources to enhance response accuracy, and optimizing computational efficiency to reduce latency. These enhancements will contribute to making RAG-based systems more effective for real-world financial analysis and decision-making.

\section*{CRediT authorship contribution statement}
\textbf{Thai Phong Huan:} Conceptualization, Methodology, Software, Writing – original draft, Writing – review \& editing. \textbf{Le Thi Anh Thu:} Methodology, Software, Writing – review \& editing. \textbf{Nhan Huu Tran :} Methodology, Software, Writing – review \& editing. \textbf{Lenh Cong Phan pham:} Methodology, Software, Writing – review \& editing. \textbf{Hong Nhung Nguyen:} Supervision, Methodology, Validation.
\begin{comment}
For transparency, we require corresponding authors to provide co-author contributions to the manuscript using the relevant CRediT roles. The CRediT taxonomy includes 14 different roles describing each contributor’s specific contribution to the scholarly output. The roles are: Conceptualization; Data curation; Formal analysis; Funding acquisition; Investigation; Methodology; Project administration; Resources; Software; Supervision; Validation; Visualization; Roles/Writing - original draft; and Writing - review & editing. Note that not all roles may apply to every manuscript, and authors may have contributed through multiple roles.
\end{comment}

\section*{Declaration of Competing Interest}
The authors declare that they have no known competing financial interests or personal relationships that could have appeared to influence the work reported in this paper.

% \section*{Acknowledgements}
% Collate acknowledgements in a separate section at the end of the article before the references and do not, therefore, include them on the title page, as a footnote to the title or otherwise. List here those individuals who provided help during the research (e.g., providing language help, writing assistance or proof reading the article, etc.).

% \section*{Data availability}
% Provide a link to your data if available.

\bibliography{ref}

@misc{b1,
  author       = {{Gemini Team} and others},
  title        = {{Gemini: A Family of Highly Capable Multimodal Models}},
  year         = {2023},
  month        = {Dec},
  eprint       = {2312.11805},
  archivePrefix = {arXiv},
  url          = {https://arxiv.org/abs/2312.11805}
}

@misc{b2,
  author       = {{OpenAI}},
  title        = {{GPT-4 Technical Report}},
  year         = {2023},
  month        = {Mar},
  eprint       = {2303.08774},
  archivePrefix = {arXiv},
  doi          = {10.48550/arxiv.2303.08774}
}

@article{b3,
  author       = {Kasneci, E. and others},
  title        = {{ChatGPT} for good? On opportunities and challenges of large language models for education},
  journal      = {Learning and Individual Differences},
  volume       = {103},
  pages        = {102274},
  year         = {2023},
  month        = {Apr},
  doi          = {10.1016/j.lindif.2023.102274}
}

@article{b4,
  author       = {Sallam, M.},
  title        = {The Utility of {ChatGPT} as an Example of Large Language Models in Healthcare Education, Research and Practice: Systematic Review on the Future Perspectives and Potential Limitations},
  journal      = {medRxiv},
  year         = {2023},
  month        = {Feb},
  doi          = {10.1101/2023.02.19.23286155}
}

@misc{b5,
  author       = {Wu, S. and others},
  title        = {{BloombergGPT}: A Large Language Model for Finance},
  year         = {2023},
  month        = {Mar},
  eprint       = {2303.17564},
  archivePrefix = {arXiv},
  doi          = {10.48550/arxiv.2303.17564}
}

@misc{b6,
  author       = {Chen, Mark I-Cheng and others},
  title        = {Evaluating Large Language Models Trained on Code},
  year         = {2021},
  month        = {Jul},
  eprint       = {2107.03374},
  archivePrefix = {arXiv},
  doi          = {10.48550/arxiv.2107.03374}
}

@inproceedings{b7,
  author       = {Franco, Muriel Figueredo and others},
  title        = {{SecBot}: A business-driven conversational agent for cybersecurity planning and management},
  booktitle    = {2020 16th international conference on network and service management (CNSM)},
  year         = {2020},
  publisher    = {IEEE}
}

@misc{b8,
  author       = {Eloundou, T. and Manning, S. and Mishkin, P. and Rock, D.},
  title        = {{GPTs} are {GPTs}: An Early Look at the Labor Market Impact Potential of Large Language Models},
  year         = {2023},
  month        = {Mar},
  eprint       = {2303.10130},
  archivePrefix = {arXiv},
  doi          = {10.48550/arxiv.2303.10130}
}

@misc{b9,
  author       = {Sun, Z.},
  title        = {A Short Survey of Viewing Large Language Models in Legal Aspect},
  year         = {2023},
  month        = {Mar},
  eprint       = {2303.09136v1},
  archivePrefix = {arXiv},
  url          = {https://arxiv.org/abs/2303.09136v1}
}

@misc{b10,
  author       = {Dale, David and others},
  title        = {Detecting and mitigating hallucinations in machine translation: Model internal workings alone do well, sentence similarity even better},
  year         = {2022},
  eprint       = {2212.08597},
  archivePrefix = {arXiv}
}

@inproceedings{b11,
  author       = {Meng, Kevin and others},
  title        = {Locating and editing factual associations in {GPT}},
  booktitle    = {Advances in neural information processing systems 35},
  pages        = {17359--17372},
  year         = {2022}
}

@misc{b12,
  author       = {Colombo, Pierre and others},
  title        = {{SaulLM-7B}: a pioneering large language model for law},
  year         = {2024},
  eprint       = {2403.03883},
  archivePrefix = {arXiv}
}

@misc{b13,
  author       = {Zhang, H. and others},
  title        = {{HuatuoGPT}, towards Taming Language Model to Be a Doctor},
  year         = {2023},
  month        = {May},
  eprint       = {2305.15075},
  archivePrefix = {arXiv},
  doi          = {10.48550/arxiv.2305.15075}
}

@misc{b14,
  author       = {Guu, K. and Lee, K. and Tung, Z. and Pasupat, P. and Chang, M.-W.},
  title        = {{REALM}: Retrieval-Augmented Language Model Pre-Training},
  year         = {2020},
  month        = {Feb},
  eprint       = {2002.08909},
  archivePrefix = {arXiv},
  url          = {https://arxiv.org/abs/2002.08909}
}

@misc{b15,
  author       = {Lewis, P. and others},
  title        = {Retrieval-Augmented Generation for Knowledge-Intensive {NLP} Tasks},
  year         = {2021},
  month        = {Apr},
  eprint       = {2005.11401},
  archivePrefix = {arXiv},
  url          = {https://arxiv.org/abs/2005.11401}
}

@misc{b16,
  author       = {Wu, S. and Xiong, Y. and Cui, Y. and Wu, H. and Chen, C. and Yuan, Y. and Huang, L. and Liu, X. and Kuo, T.W. and Guan, N. and Xue, C.J.},
  title        = {Retrieval-augmented generation for natural language processing: A survey},
  year         = {2024},
  eprint       = {2407.13193},
  archivePrefix = {arXiv}
}

@inproceedings{b17,
  author       = {Akita, R. and Yoshihara, A. and Matsubara, T. and Uehara, K.},
  title        = {Deep learning for stock prediction using numerical and textual information},
  booktitle    = {2016 IEEE/ACIS 15th International Conference on Computer and Information Science (ICIS)},
  year         = {2016},
  doi          = {10.1109/ICIS.2016.7550882}
}

@misc{b18,
  author       = {Sen, J. and Mehtab, Sidra},
  title        = {A Robust Predictive Model for Stock Price Prediction Using Deep Learning and Natural Language Processing},
  year         = {2021},
  month        = {Jul},
  doi          = {10.36227/techrxiv.15023361.v1}
}

@inproceedings{b19,
  author       = {Patel, R. and Choudhary, V. and Saxena, D. and Singh, A. K.},
  title        = {{LSTM} and {NLP} Based Forecasting Model for Stock Market Analysis},
  booktitle    = {2021 First International Conference on Advances in Computing and Future Communication Technologies (ICACFCT)},
  year         = {2021},
  month        = {Dec},
  doi          = {10.1109/icacfct53978.2021.9837384}
}

@inproceedings{b20,
  author       = {Alhydary, T. and Qashqary, L. and Alotaibi, S. and Jaha, R. and Alansari, S.},
  title        = {Smart Trader: Towards an Advisory Chatbot Service for Saudi Stock Market Investment},
  booktitle    = {2024 Seventh International Women in Data Science Conference at Prince Sultan University (WiDS PSU)},
  pages        = {32--35},
  year         = {2024},
  month        = {Mar},
  doi          = {10.1109/wids-psu61003.2024.00021}
}

@inproceedings{b21,
  author       = {Lewis, Patrick and others},
  title        = {Retrieval-augmented generation for knowledge-intensive {NLP} tasks},
  booktitle    = {Advances in Neural Information Processing Systems 33},
  pages        = {9459--9474},
  year         = {2020}
}

@misc{b22,
  author       = {Jiang, Albert Q. and others},
  title        = {Mistral 7B},
  year         = {2023},
  eprint       = {2310.06825},
  archivePrefix = {arXiv}
}

@misc{b23,
  author       = {Reid, M. and others},
  title        = {{Gemini} 1.5: Unlocking multimodal understanding across millions of tokens of context},
  year         = {2024},
  month        = {Mar},
  eprint       = {2403.05530},
  archivePrefix = {arXiv},
  doi          = {10.48550/arxiv.2403.05530}
}

@article{b24,
  author       = {Zhang, M. and Li, J.},
  title        = {A commentary of {GPT-3} in {MIT} Technology Review 2021},
  journal      = {Fundamental Research},
  volume       = {1},
  number       = {6},
  pages        = {831--833},
  year         = {2021},
  month        = {Nov},
  doi          = {10.1016/j.fmre.2021.11.011}
}

@inproceedings{b25,
  author       = {Li, Zonglin and Guo, Ruiqi and Kumar, Sanjiv},
  title        = {Decoupled Context Processing for Context Augmented Language Modeling},
  booktitle    = {Advances in Neural Information Processing Systems 35 (NeurIPS)},
  year         = {2022}
}

@misc{b26,
  author       = {Wang, Boxin and Ping, Wei and McAfee, Lawrence and Xu, Peng and Li, Bo and Shoeybi, Mohammad and Catanzaro, Bryan},
  title        = {{InstructRetro}: Instruction Tuning post Retrieval Augmented Pretraining},
  year         = {2023},
  eprint       = {2310.07713},
  archivePrefix = {arXiv}
}

@article{b27,
  author       = {Echambadi, Vivaan},
  title        = {Financial Market Sentiment Analysis Using {LLM} and {RAG}},
  journal      = {SSRN},
  year         = {2025},
  doi          = {10.2139/ssrn.4939798}
}

@misc{b28,
  author       = {Li, X. and others},
  title        = {{AlphaFin}: Benchmarking Financial Analysis with Retrieval-Augmented Stock-Chain Framework},
  year         = {2024},
  month        = {Mar},
  eprint       = {2403.12582},
  archivePrefix = {arXiv},
  url          = {https://arxiv.org/abs/2403.12582}
}

@misc{b29,
  author       = {Zhu, Fengbin and others},
  title        = {{FinTMMBench}: Benchmarking Temporal-Aware Multi-Modal {RAG} in Finance},
  year         = {2025},
  eprint       = {2503.05185},
  archivePrefix = {arXiv}
}

@misc{b31,
  author       = {Feng, F. and Yang, Y. and Cer, D. and Arivazhagan, N. and Wang, W.},
  title        = {Language-agnostic {BERT} Sentence Embedding},
  year         = {2020},
  month        = {Jan},
  eprint       = {2007.01852},
  archivePrefix = {arXiv},
  doi          = {10.48550/arxiv.2007.01852}
}

@article{b32,
  author       = {Zhang, X. and Wu, P. and Cai, J. and Wang, K.},
  title        = {A Contrastive Study of Chinese Text Segmentation Tools in Marketing Notification Texts},
  journal      = {Journal of Physics: Conference Series},
  volume       = {1302},
  number       = {2},
  pages        = {022010},
  year         = {2019},
  month        = {Aug},
  doi          = {10.1088/1742-6596/1302/2/022010}
}

@misc{b33,
  author       = {Douze, M. and others},
  title        = {The {Faiss} library},
  year         = {2024},
  month        = {Jan},
  eprint       = {2401.08281},
  archivePrefix = {arXiv},
  url          = {https://arxiv.org/abs/2401.08281}
}

@article{b34,
  author       = {Robertson, S.},
  title        = {The Probabilistic Relevance Framework: {BM25} and Beyond},
  journal      = {Foundations and Trends® in Information Retrieval},
  volume       = {3},
  number       = {4},
  pages        = {333--389},
  year         = {2010},
  doi          = {10.1561/1500000019}
}

@article{b35,
  author       = {Mondillo, Gianluca and others},
  title        = {Basal knowledge in the field of pediatric nephrology and its enhancement following specific training of {ChatGPT-4} ‘omni’ and {Gemini} 1.5 Flash},
  journal      = {Pediatric Nephrology},
  year         = {2024},
  month        = {Aug},
  doi          = {10.1007/s00467-024-06486-3}
}

\end{document}